\documentclass[12pt,a4paper]{article}

\usepackage[a4paper,margin=2cm, marginparwidth=2cm]{geometry}

\usepackage{amsmath,amssymb,amsthm}

\usepackage{hyperref}

\usepackage{graphicx}

\usepackage[dvipsnames]{xcolor} 

\usepackage[margin=10pt,font=small,labelfont=bf,labelsep=endash]{caption}

\newcommand{\tsenote}[1]{}
\newcommand{\tsecomment}[1]{}

\providecommand{\eqref}[1]{(\ref{#1})}

\newcommand{\figref}[1]{Fig.~\ref{#1}}

\newcommand{\secref}[1]{section \ref{#1}}
\newcommand{\appref}[1]{Appendix~\ref{#1}}
\newcommand{\Figref}[1]{Fig.~\ref{#1}}

\providecommand{\ref}[2]{#2\footnote{See \texttt{#2}.)}}

\newcommand{\Acal}{\mathcal{A}}
\newcommand{\Ccal}{\mathcal{C}}
\newcommand{\Hcal}{\mathcal{H}}
\newcommand{\Mcal}{\mathcal{M}}
\newcommand{\Tcal}{\mathcal{T}}
\newcommand{\Wcal}{\mathcal{W}}

\newcommand{\Wbar}{\overline{W}}

\newcommand{\BHI}{\widehat{\mathcal{W}}}

\newcommand{\Pfirst}{P^\mathrm{(1)}}
\newcommand{\Psecond}{P^\mathrm{(2)}}

\newcommand{\Frirm}{\mathrm{Fri}}
\newcommand{\Othrm}{\mathrm{Oth}}

\newcommand{\areaset}{\mathcal{A}} 

\begin{document}


\begin{center}
	{\Large\textbf{The Behavioural House Indicator:\\ 
      A faster and real time small-area indicative \\
      deprivation measure 
      for England}	
      }
	\\[0.5\baselineskip]
	{\large {
\href{https://https://www.imperial.ac.uk/people/e.viegas11}{Eduardo Viegas}\textsuperscript{1,2}, 
\href{https://https://www.imperial.ac.uk/people/t.evans}{Tim S.\ Evans}\textsuperscript{1}}
	}
	\\[0.5\baselineskip]
 \textsuperscript{1}\href{https://www.imperial.ac.uk/complexity-science}{Centre for Complexity Science}, Imperial College London, SW7 2AZ, United Kingdom
    \\
    \textsuperscript{2}Department of Computer Science, School of Computing, Tokyo Institute of Technology, Yokohama 226-8502, Japan
    \\[0.5\baselineskip]
    19th February 2024
\end{center}







\begin{abstract}

Researchers have been long preoccupied with the measuring and monitoring of economic and social deprivation at small scales, neighbourhood, level in order to provide official government agencies and policy makers with more precise data insights. Whist valuable methodologies have been developed, the exercise of data collection associated with these methods tend to be expensive, time consuming, published infrequently with significant time delays, and subject to recurring changes to methodology.
Here, we propose a novel method based on a straightforward methodology and data sources to generate a faster and real time indicator for deprivation at different scaling, small to larger areas. The results of our work show that our method provides a consistent view of deprivation across the regions of England and Wales, which are inline with the other indexes, but also highlight specific flash points of deep rural and highly dense urban deprivation areas that are not well captured by existing indexes. Our method is intended to aid researchers and policy makers by complementing existing but infrequent indexes.

\end{abstract}

\section{Introduction}
    
The need to identify, measure and categorise distinct levels of economic and social inequalities is a common issue that preoccupies both academic researchers and government agencies \cite{Depr2019, DeprT2019,GREEN201811,Lloyd2023}. To address this, a number of indices \cite{Depr2019,GREEN201811,Lloyd2023, Census} have been developed to compare poverty and deprivation in distinct regions across the land at both micro and macro levels. Unsurprisingly, each of these indexes provide distinct results and highlight specific features that are not easily captured the others, as there are distinct perspectives on exactly \emph{what} is being measured as well as \emph{how} to measure. 

In relation to \emph{what} is being measured, it can be argue that the distinction between poverty and deprivation is broadly agreed, as colloquially explained by the Office for National Statistics\cite{DeprS2019}: ``People may be considered to be living in poverty if they lack the financial resources to meet their needs, whereas people can be regarded as deprived if they lack any kind of resources, not just income''. The general understanding, therefore, effectively implies that \emph{poverty} is fundamentally about money (i.e.\ wealth, income, etc.), and a sub component of the overall concept of \emph{deprivation}. The definition of money, however, is much more vague as \emph{any kind of resource} can be measured and interpreted in many ways, and this is where each of the indices start to deviate conceptually, and so results vary significantly.

When it comes to \emph{how} to measure, the traditional approach \cite{Depr2019,GREEN201811,Lloyd2023, Census} taken by researchers is to narrow down their specific conceptual definition of deprivation by compartmentalising deprivation into dimensions, or sub components, so that it becomes easier to define each \emph{lack of resource} more specifically. Each domain is then weighted, either implicitly or explicitly, and an overall single index is produced. This is essentially the general approach adopted by all existing indexes described in this paper.

The above approach, however, requires significant level of arbitrary judgements \cite{Depr2019, DeprT2019} both in assigning the hierarchy and importance of each domain in the weightings used, and in the methods to harmonise individual measures of a monetary (financial and economic)  or non-monetary (social) nature. 
Inevitably, the need to address these issues tends to lead to extensive and complex methodologies that require continuous re-assessment every time new data is published, together with expensive and time-consuming data collection exercises.
As important is the fact that the process becomes costly and publication are results are infrequent with long time lags. It is of little surprise then that authorities and bodies such as the Office for National Statistics in the United Kingdom\cite{ONS0124} 


It is within this context that we place the aims of and motivation for our research. Fundamentally, we aimed to develop a simple measure, the Behavioural House Indicator $\BHI$, that can be used to measure \emph{deprivation} consistently and fast in short real-time intervals. Moreover, we intend to fill the existing large gaps between the timelines of publication of the other more established indexes by comparing and contrasting the results.

In order to do so, we developed a framework and methodology that it is computationally simple and does not require any sort of parameters or data transformations. Moreover, we make use of a single data source, the ``\href{https://www.gov.uk/government/organisations/land-registry/about}{HM Land Registry}''. 
The HM Land Registry is a government controlled entity that keeps the all relevant property records. In particular, each time a property is changes ownership, a new transaction record is recorded at HM Land Registry.
This data is (a) factual and objective (i.e.\ no assumptions at its construction), (b) publicly available (i.e.\ at no cost), (c) published promptly and at short time intervals (i.e.\ fast) and (d) detached to from any economic and social data collection (i.e.\ independent). In addition, the data set is longstanding, extensive, and methodologically stable.


Our process, detailed below, is to use HM Land Registry data to obtain sequences of transactions for each individual property. From this we construct a conditional probability table for the day of the week of pairs of subsequent transactions. These are studied for each region (LSOA or MSOA): we calculate the corresponding mutual information \cite{Shannon1948,Viegas2020} for each region, determine the average mutual information of the neighbours for each region, and generate our Behavioural House Indicator $\BHI$ based on the relationship between the two quantities of mutual information.

Our proposed framework and method is underpinned by some important conceptual principles derived from complexity \cite{West2017,Huan2014}, information theory \cite{Shannon1948,Sanchirico2008,Dehmer2011}, and mathematical finance \cite{Rebonato}. Fundamentally, we highlight the fact that we are dealing with economic and social complex systems that have features and elements highly dependent and intertwined to each other \cite{Kumar2023}, and therefore with high levels of correlation and causality. This fact in isolation has important implications.

Firstly, it follows that any measure for deprivation must be mathematically \emph{non-additive} so that the effects of correlations are not compounded \cite{Rebonato,ChateauneufAlain2022Trnp}. This is fundamentally distinct from the traditional \emph{additive} approach of aggregating different measures (where the weightings are effectively an attempt to correct from the lack of computation of correlations).

Secondly, in order to avoid \emph{proverty} (monetary) measures trumping the social elements, we purposefully avoid any monetary quantities. This also has the additional benefit these quantities tend to be noisy given the temporal and relative nature of money as storage of value, and we can avoid complex financial issues \cite{alma9910002980001591,alma999870150001591} such as inflation, purchase parity, non-declared earnings, an so on that have different effects in distinct economic regions.

Lastly, we rely on the general research findings that different forms and levels of \emph{deprivation} leads to a range of distinct behaviours \cite{AnandPaul2021Npad}. As we are not preoccupied with specific definitions,  
we make use of mutual information as the preferred measure to compute changes to behaviours.
Importantly, this means that our research is geared towards the relative measurement and detection of areas of high level deprivation. It does not, however, provides any specific explanation to the specific causes of deprivation.

Here, we emphasise the difference between the methodological framework and the specific mathematical method to measure the indicator. This is simply due to the fact that we believe that the methodological framework described may be applicable most economies. However, the specific method and quantification is fundamentally dependent on the structure and the legislation of the economic system in question, and fundamentally dependent on the availability of data sources. Our data provides us with information on England and Wales, two of the four main administrative divisions of the United Kingdom.

Within this context, we narrow our study to the economies of England and Wales. We exclude the two other regions of the United Kingdom, Scotland and Northern Ireland, for two reasons. Firstly, the same level of data is not publicly and freely available and, secondly, the laws and regulations concerning house transactions in Scotland are fundamentally distinct from those of England and Wales. Furthermore, the number of yearly property transactions in Scotland and Northern Ireland are much smaller, being around 10 and 3 percent, respectively, of all transactions within the United Kingdom. In some cases, when comparing our Behavioural House Indicator $\BHI$ to other measures and indices, we will further limit ourselves to data from just England to ensure we match like with like.


In order to understand how our Behavioural House Indicator $\BHI$ performs we make comparisons with other measures produced by the \href{https://www.ons.gov.uk/}{Office for National Statistics}. This is United Kingdom government agency responsible for the collection, analysis and publication of statistical data associated with the  economy, society and population of the United Kingdom. Whilst all data related to employment, education and health for England is within the Office for National Statistics remit, some of these elements related to the remaining regions of the United Kingdom (Scotland, Northern Ireland and Wales) are the responsibility of their respective devolved governments, including the all important ten-year census. For consistency therefore, some of our analysis is for England only.

The Office for National Statistics provides statistical data to government departments, and it is independent from the government, ultimately answering to the parliament of the United Kingdom through the UK Statistics Authority. However, the publication of indices as well as the underlying methodologies are the responsibility of government departments. This is an important distinction. The Office for National Statistics and the UK Statistics Authority maintain full control of the source of the data, the Census. However, one of the most important indexes, the English Index of Multiple Deprivation published by the Ministry of Housing Communities \& Local Government every three to five years, is \emph{not} independent at all from the UK Government machinery. This fact seems to be at time either misunderstood or over brushed in research literature in this area. Further details for the indices used in this research can be found within the {Methods} section, \secref{s:meth}.

The final aspect in our analysis is the definition of the geographical areas used for our analysis areas.  We will use data from the postal service, to link properties listed in HM Land Registry transactions to postcodes. Postcodes identify small regions, typically containing 15 and usually less than 100 neighbouring properties. The Office for National Statistics often provides information on three larger scales. We will discuss these definitions of geographical areas as needed in \secref{s:meth} with additional information given in the Appendix. 

It is important to note that some measures are tied to one predefined geographical structure. For example the Index for Multiple Deprivation is only given in terms of one of these definitions, the smallest scale from the Office for National Statistics known as the Lower layer Super Output Area. In this respect, our Behavioural House Indicator $\BHI$ has a clear advantage in that it is available at a fine grained level and is easily computed for any geographical structure required. 

\section{Methods} \label{s:meth}

\subsection{Data Sources and Collection}

We make use of three distinct groups of data sets based on their usage. The first group is the property transaction data, sourced from the HM Land Registry, which is used to compute our fast, real-time, Behavioural House Indicator $\BHI$ under the methodology described later. The second group consists of the geographical data sourced from the Office for National Statistics and the Ordinance Survey 
which is used to map the property postcodes to the distinct area aggregations (such as the LSOA11, LSOA21 and MSOA21 discussed in \secref{s:geodata}), to pinpoint geographical locations, to compute the density and to determine the neighbours of the relevant area aggregations. The third group relates to social and economic data sourced from the Census 2021 or index data published by government agencies or academic research. This last set of data does not form part of any input to our methodology which is totally independent from these sources. Instead, these data sources are solely used for results comparison and benchmarking and they are not transformed or modified in any way beyond simple arithmetic operations to support aggregation and the computing of basic statistical measures. 

\subsubsection{Property related data}

The core data for our analysis is the \emph{Price Paid} database sourced from the HM Land Registry for England and Wales. The data set records basic data (i.e.\ the transaction date and price paid) as well as the address for every single registered residential property sale in England and Wales from January 1995 to June 2023. Monetary data (i.e.\ the transaction value) is only used as a filter to eliminate a very few number of entries that are inconsistent with a sale under normal conditions (i.e 6,922 entries or 0.02\% of all transactions), but subsequently, the data is purposefully discarded to ensure that we do not include any monetary measure in our analysis. After the application of data filters for entries that are not related to the sale of residential property transaction (see Appendix), the resulting final number of entries for computation are \emph{27,186,352} where English and Welsh transactions account for \emph{25,894,123} and \emph{1,292,229} respectively.

We make use of the transaction date and the postcodes as raw data fields. We compute and assign an unique ID $h$ for the property by matching the address fields (flat and house numbers, street and postcode) within the data set.  

\subsubsection{Geographical data}\label{s:geodata}

Geographical data is sourced from the Office for National Statistics via the \texttt{data.gov.uk} website. These consist of the Postcode directory, data on the network of roads, and the shape files of geographical boundaries including countries, local authorities, Output Areas `OA', Lower Layer Super Output Areas (LSOAs),  and Middle Layer Super Output Areas (MSOAs). 
Some of these boundaries are redefined after a census (carried our every ten years) so we refer to `LSOA11' and `OA21' to show that these are the definitions for regions defined for the census 2011 and 2021 respectively.  In addition, auxiliary mapping files (equivalence between LSOA11 to LSOA21, for instance) are also sourced.  The LSOAs are the most granular level of analysis carried out by policy makers, and are the basis for our research. 


\subsubsection{Economic and Social Indices}

We make use of two main data sources in relation to the economic and social indices. 

The first data set we use is related to the English Index of Multiple Deprivation $\Mcal$ for 2019 \cite{Depr2019,DeprS2019,DeprT2019} published by the Ministry of Housing Communities \& Local Government and again sourced from the \href{http://data.gov.uk}{\texttt{data.gov.uk}} website. The index has an embedded hierarchical aggregation with four layers, which, from top to bottom, are referred to as: the overall index, (seven) ``domains'', ``subdomains'', and ``indicators''. To avoid misunderstanding, we note that the use of the word \emph{indicator} within  the framework of the Index of Multiple Deprivation $\Mcal$ is substantially different from our conceptual definition described in the introduction. Here, the word is simply applied to define an underlying data source that is used to calculate a subdomain. 

The second data source is that of the 2021 Census \cite{Census}, including population, households by deprivation dimensions, ethnic groups and others. This data comes from the Office for National Statistics office census and labour market statistics website \href{http://nomisweb.co.uk}{\texttt{nomisweb.co.uk}}.
For more general references, analysis and consistency purposes, we also make use of the Access to Healthy Assets and Hazards for Great Britain ('AHAH') \cite{GREEN201811}, which is part of the data available through the Consumer Data Research Centre. The multidimensional index summarises health-related features of neighbourhoods, such as the retail environment, health services, the physical environment and the air quality that have some high level of overlapping with both $\Mcal$ and the Census households by deprivation dimensions. This data contains four ``dimensions'' of deprivation that are used for computing the level of deprivation.

As detailed footnote, $\Mcal$ was published based on the existing LSOAs at the time, i.e.\ 2011, and there were a very limited number of changes between 2011 and 2021. As the Census 2021 data, is based on the latter date, we remapped $\Mcal$ to LSOA21, where a split of a region maintains the same value for both new regions and merger changes the value to the simple average numbers of the previous regions.

\subsection{Data Analysis and Construction of the Behavioural House Indicator $\BHI$}

Each property transaction $\tau \in \Tcal$ is represented by a pair $(h,t)$ where $h \in \Hcal$ is a unique property and $t$ is the date of transaction $t$ (e.g.\ our first date is 1st January 1995). The set of all property transactions $\Tcal$  contains around $27.2$ million transactions involving around $14.8$ million  distinct properties in $\Hcal$ over the period ranging from 1st January 1995 to 30th June 2023.

We are interested in the day of the week so we define a map $D(t)$ from a date $t$ to a day $\Wcal \in \{\mathrm{Monday}, \ldots,  \mathrm{Sunday} \}$.

\label{s:mb}

\subsubsection{Frequency of transactions and transaction pairing probabilities}

The frequency of transactions $\phi_{d}$ for a given day of the week $d \in \Wcal$ can be written as
\begin{equation}
    \phi(d) 
    = 
    \frac{1}{|T|} \sum_{(h,t) \in \Tcal} \delta(d,d(t))
    \label{eq:frequency}
\end{equation}
where $\delta(x,y)=1$ if $x=y$ and is zero otherwise.

\label{s:mb1}

\subsubsection{Weekday pairing probability} 

We now focus on days of the week, $W =\{\mathrm{Monday}, \ldots, \mathrm{Sunday} \}$. We are particularly interested in the potential conditional and causal relationships in the dates of transactions of the same property.

A simple way to study this is to look at consecutive transactions of one property.  
To do this we define a sales history $s(h)$ for a given property $h$ which is the sequence $s(h) = [t_1,t_2, \ldots, t_n]$ where $t_i<t_{i+1}$ and $\{ (h, t_i ) | t_i \in s(h) \}$ is the subset of all transactions involving the property $h$.

For every single transaction $(h,t_i)$ from the sales history $s(h)$ of each property $h \in \Hcal$, we find the next 
transaction for that property which will be the transaction $(h,t_{i+1})$. 
Now we can then count the number of consecutive transactions $L(d_1,d_2)$ for the same property where the first transaction happens on day $d_1=d(t_i)$ while the subsequent is on day of the week $d_2=d(t_{i+1})$
\begin{eqnarray}
    L(d_1,d_2)
    &=&
    \sum_{(h,t_i) \in \Tcal}
    \delta(d_1,d(t_i)) \,
    \delta(d_2,d(t_{i+1}))
    \, .
    \label{e:Ldef}
\end{eqnarray}
This gives us the probability of a joint transaction $P(d_1,d_2)$ as 
\begin{eqnarray}
    P(d_1,d_2) 
    &=&
    \frac{1}{Z} L(d_1,d_2) \, , \quad d_1,d_2 \in \Wcal
    \label{e:Pdoubledef}
    \\
	Z 
	&=& 
	\sum_{d_1 \in \Wcal} \sum_{d_2 \in \Wcal} L(d_1,d_2)
	\, .
\end{eqnarray}
Note that if a transaction $\tau=(h,t_i)$ has no subsequent transaction then we define $\delta(d_2,d(t_{i+1}))=0$ in such cases. That is the set of transactions which give non-zero contributions is $\Tcal^*$ which only contains transactions of properties that are sold at least twice so $\Tcal^* \subset \Tcal$.

The single probability distributions $\Pfirst$ and $\Psecond$ are the probabilities that a transaction chosen uniformly at random from the reduced set of transactions $\Tcal^*$ (those which involve properties with at least two transactions in the full data $\Tcal$) is 
\begin{eqnarray}
    \Pfirst(d_1) &=& \sum_{d_2\in \Wcal} \frac{L(d_1,d_2)}{Z}  = \sum_{d_2\in \Wcal} P(d_1,d_2)
    \, , 
    \label{e:Pfirstdef}
    \\
    \Psecond(d_2) &=& \sum_{d_1\in \Wcal} \frac{L(d_1,d_2)}{Z}  = \sum_{d_1\in \Wcal} P(d_1,d_2)
    \label{e:Pseconddef}
    \, .
\end{eqnarray}

\label{s:mb2}

\subsubsection{Mutual Information}\label{s:mb3}

The pointwise contribution to the mutual information $I(d_1,d_2)$ and the total mutual information $W$ \cite{Shannon1948,Viegas2020} is therefore:
\begin{eqnarray}
    I(d_1,d_2)
    &=&
    P(d_1,d_2) \log_{2} \left( \frac{P(d_1,d_2)}{\Pfirst(d_1)\Psecond(d_2)}  \right)
    \, , 
    \label{e:Idef}
    \\
    W 
    &=& 
    \sum_{d_1 \in \Wcal} \sum_{d_2 \in \Wcal} I(d_1,d_2)
    \label{e:Wdef}
\end{eqnarray}

\subsubsection{Computing neighbourhood and the average mutual information of neighbours}\label{s:neig}

We will also examine our results for properties in specific areas. To do this start with a partition of the total area into a set of non-overlapping smaller areas $\areaset = \{a_{1}, a_{2}, \ldots, \}$. A property can only be in one of these subdomains so we denote the set of properties in area $a \in \areaset$ as $\Hcal_{a} \subset \Hcal$ and the set of transactions of properties in area ${a}$ as $\Tcal_{a}$. We can then repeat the analysis above for transactions of properties in area ${a}$, starting from the count $L$ of \eqref{e:Ldef} which becomes $L_{a}$ for transactions in area ${a}$ where
now $L_a(d_1,d_2) =	\sum_{(h,t_i) \in \Tcal_{a}} 	\delta(d_1,d(t_i)) \delta(d_2,d(t_{i+1}))$. Replacing $L$ by $L_{a}$ leads to area specific versions of quantities in \eqref{e:Pdoubledef}, \eqref{e:Pfirstdef}, \eqref{e:Pseconddef}, \eqref{e:Idef} and \eqref{e:Wdef}, 
denoted as $P_{a}(d_1,d_2)$, $\Pfirst_{a}(d_1)$, $\Psecond_{a}(d_2)$, $I_a(d_1,d_2)$ and $W_{a}$  respectively (see \appref{a:nmi} for detailed forms).


In many situations the areas we use are very small leading to large fluctuations. If we assume many neighbouring areas have similar properties, it makes sense to smooth our measures over slightly larger regions. We adopt a simple local aggregation procedure where we average over the values from an area and its neighbours. 

We define two areas ${a}$ and ${b}$ to be neighbours if any road (links or nodes) recorded within the \href{https://www.ordnancesurvey.co.uk/products/os-open-roads}{Ordinance Survey Open Road data set} crosses or touch the boundaries of both areas. We emphasise here that sharing boundaries is not a sufficient condition for $a$ and $b$ to be neighbours. Instead at least one single road must cross (or at the extreme, must touch) the boundary between these two areas. 
We encode this information in an adjacency matrix $E_{ab}$ which is $1$ if areas $a$ and $b$ are neighbours, while it is zero otherwise (including when $a=b$).
This allows us to compute a separate mutual information of an area $a$ in terms of an average of its neighbours as
\begin{equation}
    \Wbar_a 
    = \frac{1}{k_{a}} 
    \sum_{{b} \in \areaset}
    W_b E_{ab}
    \quad \text{where}
    \quad 
    k_{a} = \sum_{{b} \in \areaset} E_{ab} \, .
    \label{e:Wbardef}
\end{equation}

\subsubsection{Generating the Behavioural House Indicator $\BHI$}\label{sec:five}

A scatter plot of the mutual information $W_a$  against average neighbour mutual information $\Wbar_a$ for each area ${a} \in {A}$ we observe, as expected, an approximate linear relationship $\Wbar = \alpha W + \beta$ where $\alpha$ and $\beta$ are independent of the area and can be found from a best fit procedure.

Once these parameters have been obtained we can now use them to produce an aggregation of the mutual information of an area and its neighbours to produce our Behavioural House Indicator $\BHI_a$ for each area ${a}$. We do this by finding the point $(\BHI_a,\widehat{Y}_a)$ on the best fit line $\Wbar = \alpha W + \beta$ which is the closest (orthogonal projection) to the point $(W_a,\Wbar_a)$ through the equations
\begin{eqnarray}
    \check{\beta}
    &=&
    \frac{W_a}{\alpha} + \Wbar_a
    \,
    \label{e:betadef}
    \\
    \BHI_a 
    &=& 
    \frac{\check{\beta} - \beta}{\alpha+\frac{1}{\alpha}} \, .
    \label{e:Whatdef}
\end{eqnarray}


\subsubsection{Ranking areas and bins}\label{sec:six}

\tsecomment{I do not think we need to describe deciles in the main text.  When we take a rating value $\Theta$ (index, measure etc), splitting the areas into ten groups by the ranking of the areas by that rating value groups is pretty standard. Where it gets complicated is where later we split into deciles but compare values of areas in the same deciles (most deprived) but created from different rating values: $\BHI$, $\Ccal$, $\Mcal$.}

With so many areas to study, it is often helpful when discussing and visualising trends to study groups of areas with similar values of some attribute. For instance, for some real valued rating $\Theta$ (some index or measure) we often split the areas into deciles, ten equal-size subsets $\Acal(r,\Theta)$ where $r=1,2,\ldots 10$. Any area $a$ in  $\Acal(r,\Theta)$ will be in the $r$-th decile when ranked by rating value $\Theta$, with $r=1$ representing 10\% of areas with the lowest values of $\Theta$. 

%
%
%

\section{Results}

\subsection{Data Analysis: The index foundations}

Our Behavioural House Indicator $\BHI$ is built from the temporal patterns of consecutive 
house sales and purchases so a key feature is that BHI does not involve and monetary values or measures. Our method uses mutual information between prior and subsequent sales to capture the deviations in the expected patterns.

\begin{figure}[htb!]
  \begin{center}
    \includegraphics[width=1.0\textwidth]{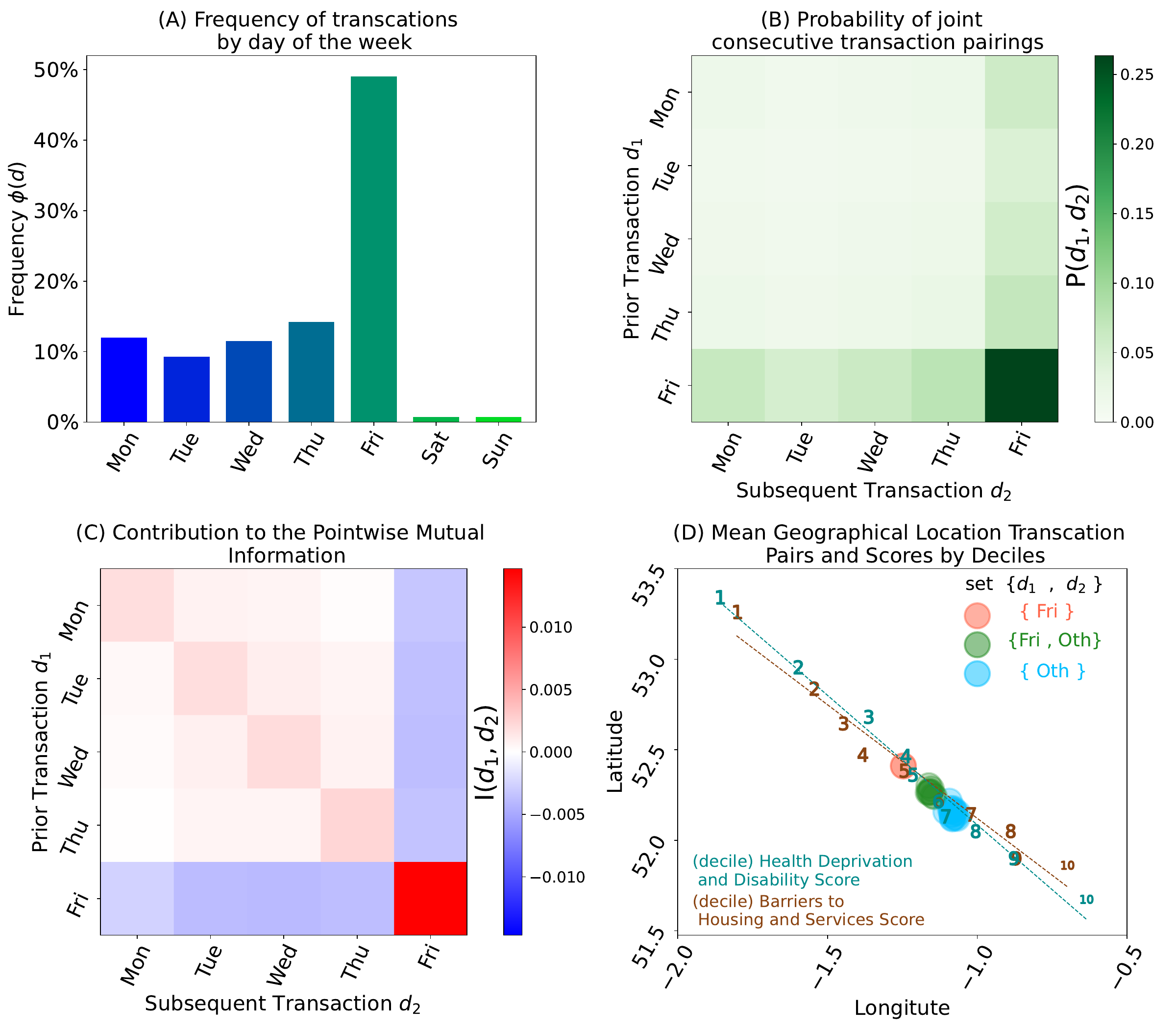}
    \caption{ \textbf{Frequency of transactions by day of the week, probabilities and contribution to the point-wise mutual information for days of the week pairing combinations and geographical interconnections.}  
    \footnotesize
      Panel (\textbf{A}) shows the frequency $\phi$ of single transactions by day of the week $(d)$, with Friday accounting for 49.8\% of all transactions. 
      Panels (\textbf{B})and (\textbf{C})relates to the combinations arising from the pairing of consecutive transaction for the same property by their respective day of the week $(d_{1},d_{2})$, where the vertical axes correspond to the prior transaction and the horizontal axes equate to the subsequent transaction. Panel (\textbf{B}) is the heat-map for the probability of pairings $P(d_{1},d_{2})$, where the combination Friday-Friday accounts for 26.6\% of all transactions.  Panel (\textbf{C}) shows the contribution to the point-wise mutual information $I(d_{1},d_{2})$ for each pairing.
      Panel (\textbf{D}) provides a geographical perspective of the interrelationship between the selected quantities based on the computation of mean coordinates (latitude and longitude). The circles represent the average coordinates of all pairings $(d_{1},d_{2})$ for three subsets of mutual information i.e.\ both on Fridays $\{ Fri \}$, Friday and another day - or vice versa - $\{ Fri , Oth \}$ and both other days $\{ Oth \}$. The lines and numbers represent the best fit and corresponding mean geographical coordinates of the deciles for each of the quantities associated with the two selected deprivation domains, namely the \emph{`Health Deprivation and Disability'} and \emph{`Barriers to Housing and Services'} scores.
      All data relates to the range between [1-Jan-1995, 30-Jun-2023]. 
      \label{fig:Data_Analysis}}
  \end{center}
\end{figure}

The relevance of the day of the week and associated measures are illustrated by the four panels in \figref{fig:Data_Analysis}, where the unique characteristics associated with Friday related house purchases can be identified. The first panel \ref{fig:Data_Analysis}A is a basic frequency bar chart that shows that on average most transactions (49.8\%) have occurred on Friday over the last 28 years. A richer insight, however, is obtained when combining the parings of days of the week between prior and subsequent dates of transactions (or purchase and sale of the property by the homeowner). Unsurprisingly, panel \ref{fig:Data_Analysis}B shows the prevalence of the Friday-Friday pairing (26.6\%), but most importantly, from the perspective of a single homeowner, the date of purchase (i.e.\ prior transaction) conditionally affects the date of sale (i.e.\ subsequent transaction). The contribution to the point-wise mutual information $I(d_1,d_2)$ is shown panel \figref{fig:Data_Analysis}C and it indicates that the Friday-Friday combination occurs much more often than would be expected from random pairs of transactions ($I(\Frirm,\Frirm) > 0$). As a result the combination of a transaction on Friday with a previous or subsequent transactions on another day occurs less often than we would expect for random independent house purchases, $I(\Frirm, \Othrm)<0$ and $I(\Othrm, \Frirm)<0$. 

Panel \ref{fig:Data_Analysis}D is an illustration of the important association and inter-dependencies concerning the geographical location of the transactions. The mean geographical coordinates (based on latitude and longitude of the postcodes) for transactions in one of three categories are shown as coloured circles. The three types of transaction pairs shown are: both Fridays (${\Frirm}$), Fridays and another day in either order ($\{\Frirm,\Othrm\}$), and both on days other than Friday (${\Othrm}$). 

On the same panel \ref{fig:Data_Analysis}D, we show the property transactions split into deciles by for the scores of a selected deprivation domain. We do this by mapping postcodes to LSOA11) to the corresponding deciles (determined in accordance with the process described in \secref{sec:six}). We show the results for two deprivation domains: the \emph{`Health Deprivation and Disability'} in blue and the \emph{`Barriers to Housing and Services'} in brown. In each case the data for a decile is shown by a number in the relevant colour with `\textbf{1}' representing the most deprived and `\textbf{10}' the least deprived by the associated domain.
As shown by the best fit lines, the dotted lines in the appropriate colour, the point-wise mutual information for these subset all lie close to these linear fits and there is a clear geographical correlation with the measures shown. This illustrates the so called `North-South divide' within England with large conurbations in the Midlands and North West often linked to post-industrial activity while the South East of England is often seen as dominated by a service-driven economy.  

\subsection{Relationship between the Behavioural House Indicator $\BHI$ and other deprivation indices}

In this section we highlight the relationship between our Behavioural House Indicator $\BHI$ and two widely used deprivation related indexes and data sets, namely the English Index of Multiple Deprivation $\Mcal$ and its seven domains, and the Census 2021 `Households by deprivation dimensions' data $\Ccal$.

\subsubsection{The English Index of Multiple Deprivation $\Mcal$}

\begin{figure}[htb!]
  \begin{center}
    \includegraphics[width=0.9\textwidth]{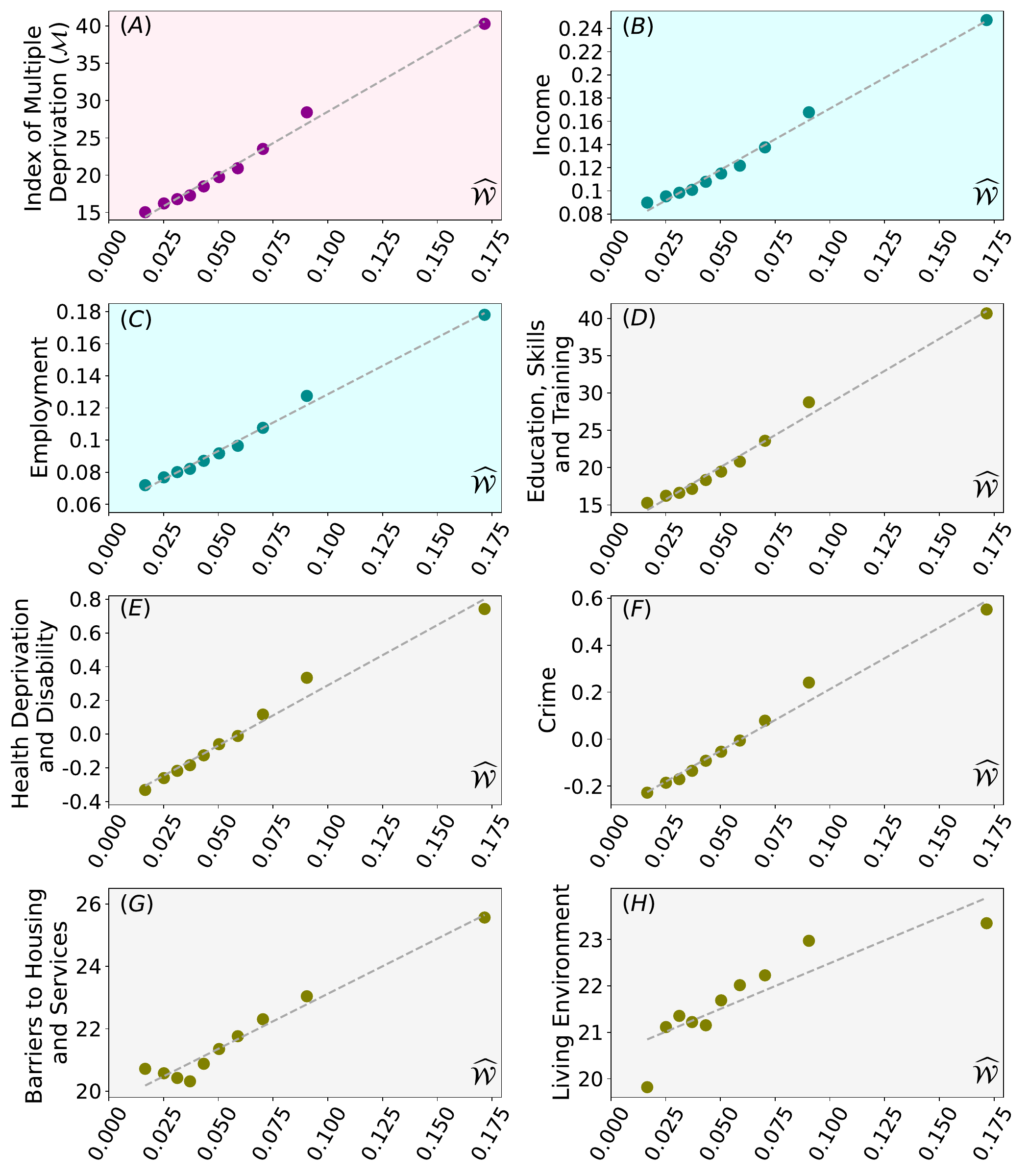}
    \caption{ \textbf{Relationship between the Behavioural House Indicator $\BHI$ and the English Index of Multiple Deprivation $\Mcal$.}  
    \footnotesize
      Each data-point corresponds to all LSOA11s within the first to the tenth deciles as a function of the ranking of $\BHI$. The x-axes are the same for all panels, corresponding to the mean scores of $\BHI$ for each data-point. In a similar manner, the y-axes equate to mean scores for the overall $\Mcal$ (panel \textbf{A}) and its seven domains (panels \textbf{B} to \textbf{H}) as labelled. The dotted grey line represent the corresponding best fit between the associated quantities. 
      \label{fig:ESIMD}}
  \end{center}
\end{figure}

The results in \figref{fig:ESIMD} show that there is a significant and meaningful relationship between $\BHI$ and the English Index of Multiple Deprivation $\Mcal$ for the year 2019 (the last available date) both at overall and at its seven domain levels. All panels within \figref{fig:ESIMD} imply high correlation levels, and good linear fitting for the data (which is aggregated into deciles and LSOA11 regions in accordance with the method described in \secref{sec:six}). The linear fitting is broken, however, at the early deciles for the domains \emph{`Barriers to Housing and Services'} and \emph{`Living Environment'}. 
On investigation of this breakdown of the subdomains and indicators related to each, we observe that the main cause of the breakdown for the former are the indicators within the \emph{`Wider Barriers to Housing and Services'} subdomain, whereas the latter is impacted by the \emph{`Indoors'} subdomain.


In order to substantiate the robustness of the relationship, in particular in relation to correlation and linear fitting, we also tested different combinations of bin sizes (not just the ten equal sized bins mentioned in \secref{sec:six}). We found that correlation levels are consistently high (unsurprisingly better when bin sizes and minimum data points increase) and the slope of the linear fitting remains largely stable regardless of the way we aggregate the data in plots of the type shown in \figref{fig:ESIMD}. 

In addition, it is important to emphasise that these results also indicate a high correlation level among the seven domains in $\Mcal$ which poses a question whether the overall methodology for computing $\Mcal$ may be simplified by relying of a much smaller set of indicators. nm

\subsubsection{The Census 2021 `Households by deprivation' dimensions $\Ccal$}

\begin{figure}[htb!]
  \begin{center}
    \includegraphics[width=0.9\hsize]{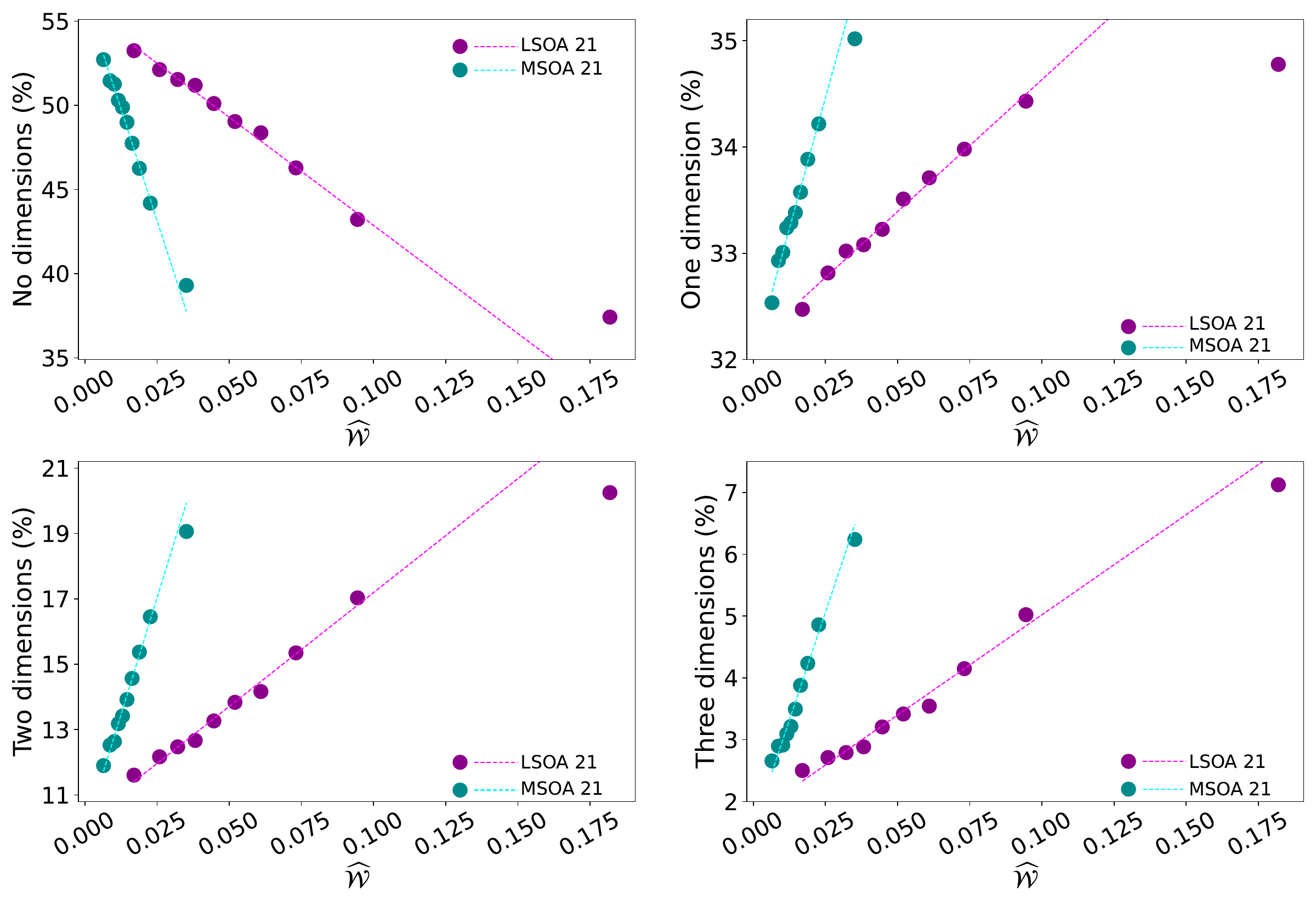}
    \caption{ \textbf{Relationship between the Behavioural House Indicator $\BHI$ and the Census 2021 households by deprivation dimensions.}  
    \footnotesize
      Each data-point corresponds to all output areas (expressed either as LSOA21 or MSOA21) within the first to the tenth deciles as a function of the ranking of $\BHI$. The x-axes are the same for all panels, corresponding to the mean scores of $\BHI$ for each data-point. In a similar manner, the y-axes equate to mean of the ratio (i.e.\ number of deprived households (in none \textbf{A}, one \textbf{B}, two \textbf{C} or three \textbf{D} of out four dimensions over the total number of households) for the output areas within the decile. The dotted grey line represent the best fit between the associated quantities, excluding the last point.  
      \label{fig:Census}}
  \end{center}
\end{figure}

As it can be observed in \figref{fig:Census}, and similarly to the English Index of Multiple Deprivation $\Mcal$, our results show that the Census 2021 data $\Ccal$ on households by deprivation dimensions have strong relationship, both in terms of correlation levels and linear fitting. Importantly, these relationships are maintained regardless of the number of dimensions (the data contains four ``dimensions'' of deprivation) that are used for computing the level of deprivation, or the size of the region of aggregation (i.e.\ LSOA21 as well as MSOA21). 

It is also important to note that $\BHI$ is a relative measure that reduces as regions are aggregated from LSOA21 to MSOA21. Yet, the ability to rank from least to most deprived is still maintained. Also, the relative reduction in value is an expected feature since variances from expectations (which is essentially the nature of mutual information) reduce as the number of properties within each region increases.  

Whist the Census does not explicitly provide an index to deprivation, one can be easily derived by the ratio between the number of households with deprivation the total number of households. For the purposes of this research, we compute $\Ccal$ based on such ratio, by making use of households with \emph{any} deprivation domain as the numerator.

\subsection{Analysis of the most deprived LSOAs}

\begin{figure}[htb!]
  \begin{center}
    \includegraphics[width=0.82\textwidth]{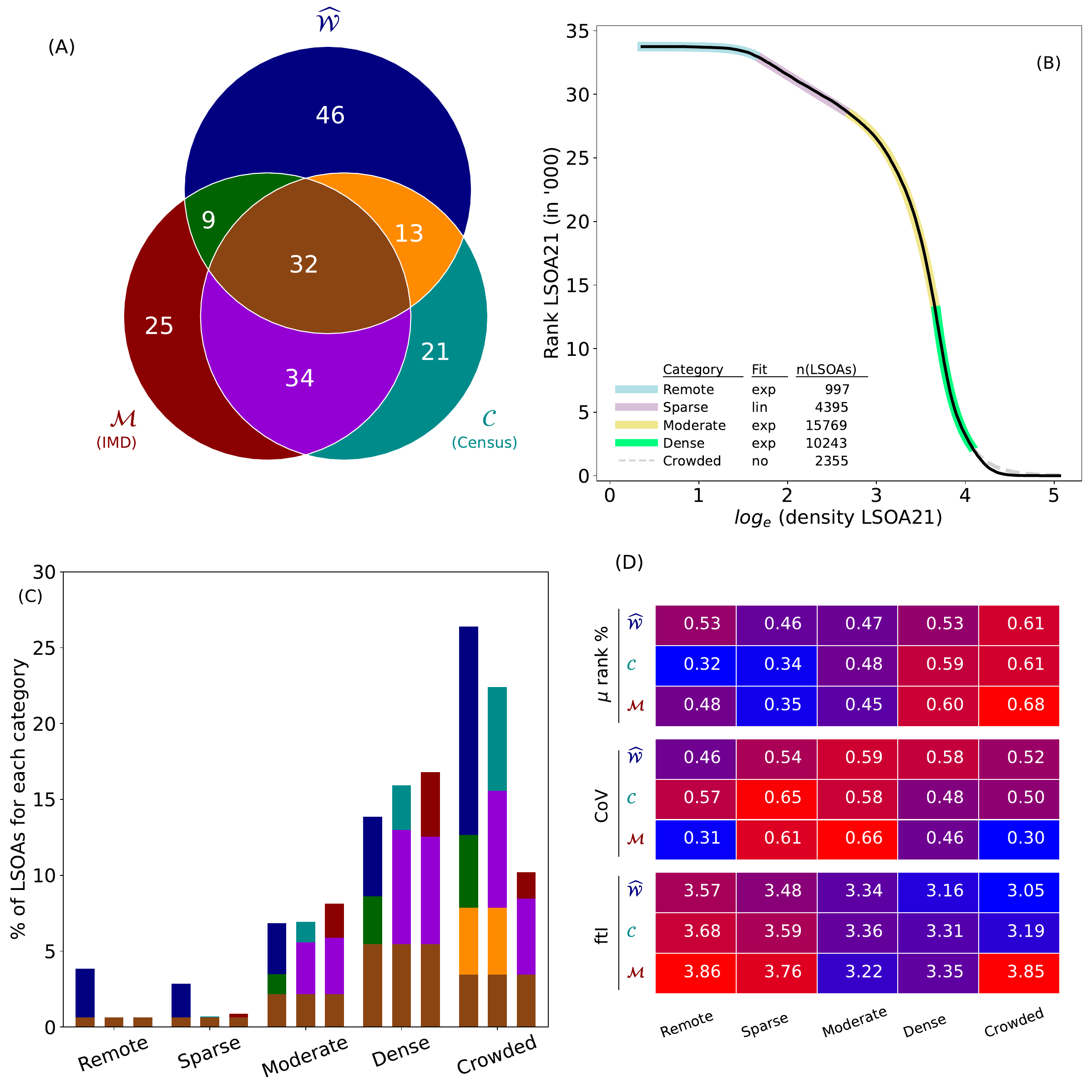}
    \caption{ \textbf{Relationship among most deprived deciles by the density categories of LSOAs.}  
    \footnotesize
      Panel \textbf{A} is a Venn diagram representation of the logical relation between the Behavioural House Indicator $\BHI$, the Census 2021 Deprivation $\Ccal$ and the English Index of Multiple Deprivation $\Mcal$ sets. Each circle represents the set of LSOAs within the most deprived decile for the corresponding deprivation measure. Numbers in one of the distinct regions of the Venn diagram reflect the percentage of the LSOAs (from any of the three measures with a non-zero contribution to that region) which lie in that intersection. For instance the dark green region represents the 9\% of LSOAs in the lowest decile BHI $\BHI$ set that are the same LSOAs which are in the lowest decile IMD $\Mcal$ set (the percentage is the same as all deciles have the same number of LSOAs). By implication, none of these LSOAs in the dark green area are in the lowest decile of the Census $\Ccal$ measure.
      Panel \textbf{B} is a semi-log plot where LSOAs are ranked from highest to least density, with the x-axis corresponding to the natural logarithm of the density and the y-axis the corresponding ranking. The thin black line corresponds to the actual data, whereas each of the coloured thick lines represent the sections with distinct fit functions and parameters where the best fit for the curve as whole is optimised. The gray dotted line is the continuation of the green line fitting at the stage that divergences increase (i.e.\ the fat tail) \cite{CohenJoelE2020Hdck}. The table within panel B provides a label for each optimised section, together with the best fitting function and the corresponding number of LSOAs with the sector. 
      Panel \textbf{C} is a cumulative histogram showing the ratio between the number of LSOAs within the most deprived decile to the total number of LSOAs within each density category. The colours correspond to the subsets in the Venn diagram of panel A. Within each density category the histograms are ordered with $\BHI$ on the left, $\Ccal$ centre, and  $\Mcal$ on the right. 
      Panel \textbf{D} corresponds to separate heat-maps (blue-purple-red, lowest to highest) for three distinct statistical measures (for the rankings of the LSOAs within each density category), from top to bottom:  the mean ranking ($\mu$ rank \%) expressed as a percentual of the total number of LSOAs , the coefficient of variance (CoV) and the fat tail level (ftl) computed as the ratio between the $97.5$ and $2.5$ centile of the ordered ranking values.   
      \label{fig:Compare}}
  \end{center}
\end{figure}

\Figref{fig:Compare}A shows there is a very good overlap of the most deprived LSOAs as measured by the three deprivation measures $\BHI$, $\Ccal$ and  $\Mcal$. This result in isolation is already remarkable given that, to be fully computed, $\Mcal$ is an index that relies on a host of detailed quantities (more than thirty variables across seven dimensions that are further subdivided into sub-dimensions), complex transformation methodologies and arbitrary weightings. In contrast, $\BHI$ is based on a single and objectively defined measure (i.e.\ the day of the week of a house purchase) based on a methodology where no weighting or parameters are required. In a similar vein, $\Ccal$ also is based on multiple data related to four separate domains (i.e.\ employment, education, health and housing) which are costly to obtain (i.e.\ once every ten years). In contrast to $\Mcal$, however, the methodology in $\Ccal$ is much simpler and there is no reliance on transforming of data and weightings (beyond the implied assumption that each dimension is equal to another). 

One can assert that the reason for the overlapping results to be so robust is the fact that there is a high level of correlation between the economic, social and environmental variables used by $\Mcal$ and $\Ccal$, and that these variables are also highly correlated to the behaviour expressed by the process of house purchases.

Furthermore, additional insight and understanding can be obtained by the detailed analysis of the \emph{differences} among the $\BHI$, $\Ccal$ and  $\Mcal$ most deprived subsets. Firstly, we adopted a data driven approach to separate LSOAs into distinct categories \cite{GartnerAndrea2011Rmdi} based on their ranking distribution of densities as shown in \figref{fig:Compare}B. We identified distinct regions within the curve of density rank against log density by optimising linear and exponential fittings through a very large number of sector combinations. We obtained five classes of density that we described from least to highest as ``remote'', ``sparse'', ``moderate'', ``dense'' and ``crowded''. Clearly, one can find a significant overlap between the more traditional Census categorisation of built up areas as well as rural and urban areas to our method. However, we opted to use this bottom-up data driven approach in order to highlight the specific features of the extreme ``remote'' and ``crowded'' LSOAs without resorting to additional data. 

Through this categorisation, we can make the following observations. Firstly, $\Mcal$ is significantly biased towards moderate and dense LSOAs as observed in \figref{fig:Compare}\textbf{C}. This effectively means that rural \cite{GartnerAndrea2011Rmdi,Kuttler} highly deprived LSOAs are almost non-existent in $\Mcal$. In addition, significant pockets of inner London that are traditionally viewed as highly deprived tend to be excluded \cite{ClokePaul1995Dpam,Lloyd2023}. This is one of the primary reasons that recent research identified $\Mcal$ to under select LSOAs with high level of ethnic minorities \cite{Lloyd2023} as there is a significant number of these minorities in the highly dense, crowded, regions. Moreover, \figref{fig:Compare}\textbf{D} shows the average rank of $\Mcal$ to significantly vary as a function of the geographical density ($\mu$ rank \%), and that the both ends of the density spectrum $\Mcal$ will tend to have low  (CoV) and heavy tails (the fat tail level is the ratio between the $97.5$ and $2.5$ centile of the values).    
In contrast, $\Ccal$ does select a significant number of crowded LSOAs, but it fundamentally tends to disregard extreme rural deprivation. Indeed, \figref{fig:Compare}\textbf{D} ($\mu$ rank \%) expresses a monotonic behaviour which is a clear indication of dependency on density. Lastly, $\BHI$ tends to be more balanced across the density spectrum, even though it shows some moderate level of bias towards the most crowded LSOAs.  

\subsection{Analysis of the remote and sparse areas}

\begin{figure}[htb!]
  \begin{center}
    \includegraphics[width=0.83\textwidth]{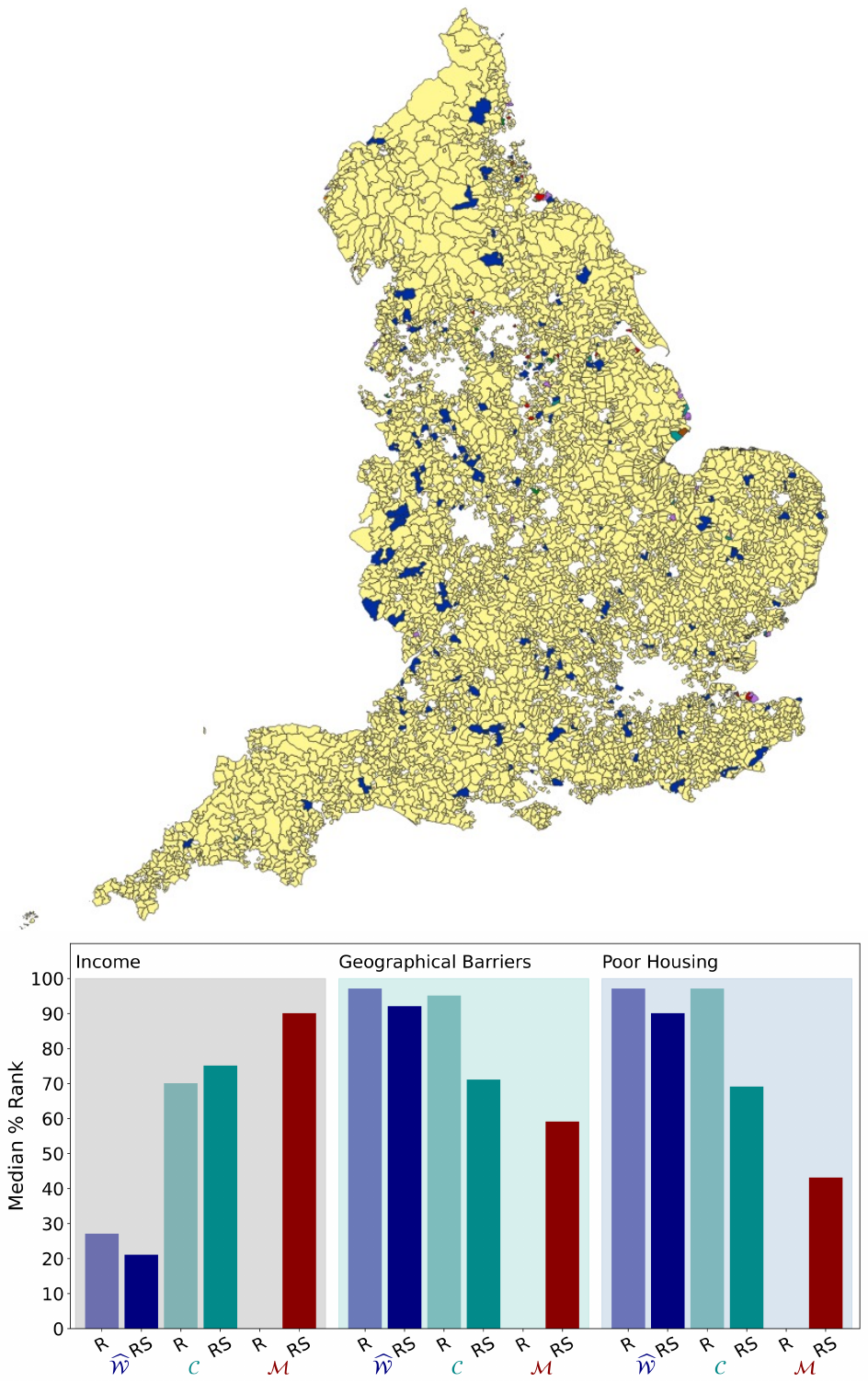}
    \caption{ \textbf{Geographical distribution of $\BHI$, $\Ccal$ and $\Mcal$ across England for the remote and sparse LSOA areas and analysis of selected deprivation indicators affecting these categories.}   
    \footnotesize
      The map at the top shows all remote and sparse areas within England. The light yellow shapes represent LSOAs that are not within the most deprived decile by any of the indexes. Dark blue , dark cyan and dark red represent the most deprived areas for $\BHI$, $\Ccal$ and $\Mcal$, respectively.
      The histogram at the bottom show the centile position of the ranking for the median of the Income (domain), Geographical Barriers (subdomain) and Poor Housing (indicator) measures that can be found within the construction of $\Mcal$. The darker shades relate to both remote and sparse regions, and lighter shades contain only the remote ones. Blue, cyan and red colour tones represent $\BHI$, $\Ccal$ and $\Mcal$, respectively.
      \label{fig:Rural}}
  \end{center}
\end{figure}

The remote and sparse areas account for 16\% of the total number of LSOAs but 87\% of the total area of England. Given these numbers, it is highly surprising that both $\Ccal$ and  $\Mcal$ rarely register a highly deprived area within these categories. Indeed, we would argue here that this analysis poses significant questions as to whether these indexes are fit for the purpose of capturing rural deprivation, as it is not unreasonable to speculate that the nature of rural deprivation is very distinct from that of urban deprivation \cite{GartnerAndrea2011Rmdi,Kuttler}. As observed within the map in \figref{fig:Rural}, $\BHI$ captures highly deprived LSOAs in a widely distributed manner across the whole country, with some pockets in Wiltshire, Herefordshire and Shropshire. In contrast $\Ccal$ and  $\Mcal$ are limited to very few coastal locations and near city areas. 

The existing bias against remote and sparse areas can be explained by the fact that $\Mcal$ is primarily driven by the highly correlated measures of income and employment (with a combined weighting of 63\%), and $\Ccal$ also has a 50\% implied weighting to the same type of dimensions. Essentially, this means that areas that are significantly impact by a more social perspective, such as those areas with high geographical barriers and extremely poor housing, are not captured as highly deprived, as the social components are diluted by primary economic and monetary dimensions. In contrast, the social component seem to affect house purchase behaviour, and therefore these are indirectly captured by $\BHI$. The histograms in \figref{fig:Rural} illustrate these observations. The very few LSOAs categorised as highly deprived by $\Mcal$ are those that have very high rank of income deprivation, but not higher ranking levels for the Geographical Barriers subdomain and the Poor Housing indicator even though both of them are part of the construction of the $\Ccal$ and the $\Mcal$ indexes. In contrast, LSOAs within remote and sparse areas that exhibit extreme high levels of geographical barriers and poor housing are captured and classified as highly deprived by $\BHI$ even if the average monetary and economic features are positive.   

It is important to emphasise that we are not judging whether one index performs better than another, as this is a relative element and it will depend on usage. Here, we are solely emphasising the positive and negative bias of each index, and the impact to different areas of social and economic structure.

\subsection{Analysis of the crowded and dense categories}

\begin{figure}[htb!]
  \begin{center}
    \includegraphics[width=0.9\hsize]{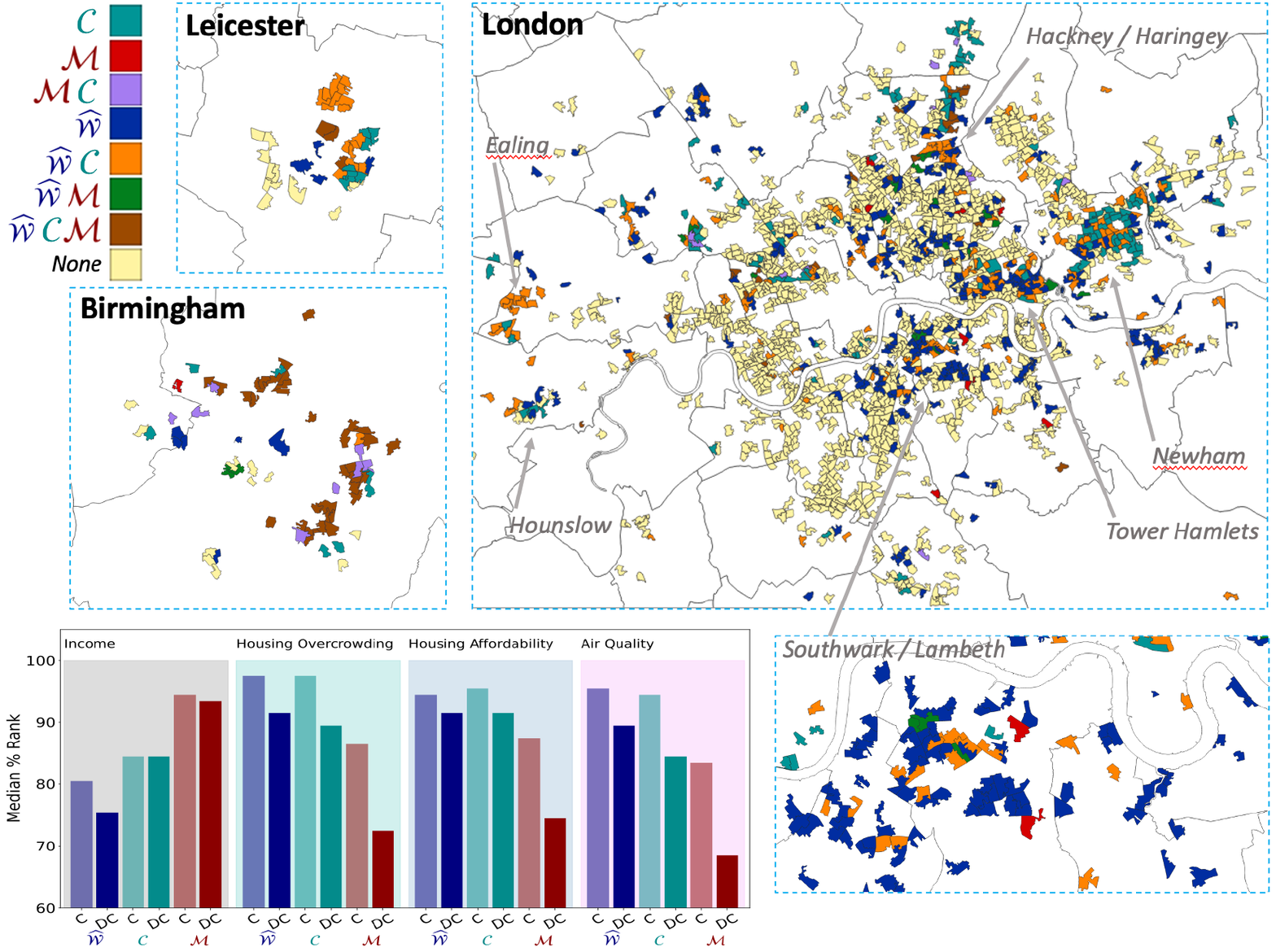}
    \caption{ \textbf{Geographical distribution of $\BHI$, $\Ccal$ and $\Mcal$ for the crowded category in selected urban areas  and analysis of selected deprivation indicators affecting dense and crowded regions.}  
    \footnotesize
      Each of the three maps at the top show the distribution of the crowded areas within London (top right), Leicester (to left) and Birmingham (middle left). The colour representations for each index $\BHI$, $\Ccal$ and $\Mcal$ and their combinations are shown at the scheme on the top left. The map at the bottom right corresponds to zoomed in cut for the London central areas at the south of the river Thames, and it excludes the LSOA crowded areas that are not regarded as highly deprived by any measure (light yellow in the other maps). 
      The histogram at bottom left shows the centile position of the ranking for the median of the Income (domain), Housing Overcrowding (indicator), Housing Affordability (indicator) and Air quality (indicator0 measures that can be found within the construction of $\Mcal$. The darker shades relate to both dense and crowded regions, and lighter shades contain only the crowded ones. Blue, cyan and red colour tones represent $\BHI$, $\Ccal$ and $\Mcal$, respectively.
      \label{fig:Urban}}
  \end{center}
\end{figure}

Both our Behavioral House Indicator $\BHI$ and the 2021 Census data $\Ccal$ show a tendency to capture higher numbers of highly deprived crowded areas, whereas English Index of Multiple Deprivation $\Mcal$ tends to manifest the reverse. This feature is well illustrated by the maps in \figref{fig:Urban} for the cities of London, Birmingham and Leicester. Whereas $\BHI$ and $\Ccal$  are well represented, and have high levels of overlapping in all these cities, $\Mcal$ is only significantly present in Birmingham (where it significantly overlaps with the others). 

Similarly to the analysis for remote and sparse regions, $\Mcal$ existing bias against dense and crowded areas can be explained by the fact that the index is primarily driven by low income and unemployment. As shown by the histogram within \figref{fig:Urban} areas with significant high levels of House Overcrowding, Housing (Low) Affordability and (Bad) Air Quality are not captured as highly deprived by $\Mcal$, and are diluted by primary economic and monetary measures that do not take into account the relative perspective of money and quality of employment. In contrast, both $\BHI$ and $\Ccal$ capture these features very well within crowded regions. 

Here, we highlight the fundamental issue that tends to arise from monetary and economic measures such as the income and employment which is best encapsulated by the expression \emph{struggling to make ends meet}. The population may be employed and have proportionally higher levels of income, and yet these factors are not enough to provide for basic needs as affordability (where housing plays a fundamental part) is extremely low. Therefore, we would argue that $\Mcal$ does not provide enough weight to deprivation indicators in urban areas where the quality of employment and relative purchase power of earning have more complex facets.

Comparing and contrasting $\BHI$ and $\Ccal$ also provides interesting additional insights. Firstly, we have already highlighted the high levels of overlap in cities such as Birmingham and Leicester as well as London, where the inner boroughs of Tower Hamlets, Newham, Hackney and Haringey, traditionally associated with high levels of deprivation, being well captured by both methods. Significant divergence between $\BHI$ and $\Ccal$, however, can be founded at the crowded LSOAs on the south bank of the river Thames (primarily the boroughs of Lambeth and Southwark) as observed in the zoomed map at the bottom right of \figref{fig:Urban}. On the south bank, $\BHI$ has a much wider LSOA representation, covering areas that are normally associated with high deprivation levels, more specially areas that were identified as initial flash points during the 2011 London riots.

\section{Discussion}

The observations made in our \emph{Results} section above indicate that our proposed Behavioural House Indicator $\BHI$ can be reasonably used as a faster, real-time, simple measure for deprivation in England and Wales. Importantly, it can be feasibly be published on a monthly basis (albeit we would argue a annual or semi-annual is sufficient as changes to deprivation are slow) with very short time delay as the computation of $\BHI$ is solely based on a single and reliable, precise, factual and objective  available data source from the UK Government which is updated at very frequent intervals. In contrast the 2021 Census data $\Ccal$ and the English Index of Multiple Deprivation $\Mcal$ are indexes that costly and expensive and published very sporadically. The Census follows a ten year cycle, whereas $\Mcal$ is dependent on specific government  with the next only planned for 2025 (with the previous ones being 2019 and 2015).  as it relies on specific data gathering processes, and a number of re-calibrations and readjustments. Within this context, we would argue that  $\BHI$ is an ideal candidate to be added as a social barometer to the Office for National Statistics list of faster indicators of economic activity. 

Importantly, our results do not advocate for the replacement of any of the indexes above. Instead of  we are able to show that each index performs well in different aspects of deprivation, a fact extensively researched in other recent papers \cite{NormanPD2015Tcgo,LloydC.D.2023DiE1}. We would argue that it is the combination of all these index that can provide a much richer picture both the levels of deprivation as well as the nature of deprivation. Indeed, $\BHI$ could be feasibly \emph{`calibrated'} to the outcomes of $\Mcal$, and be used as indicator of the latter between periods of publication. 

This is an important assertion given the existing speculation on scraping the census in 2031, and relying instead on a network of disparate public sector sources of data. Whereas our research show that there is much to gain from making use of public sector sources of data as proxy to the existing economic and social measures, the authors do not agree with any potential initiative for entirely scrapping of the Census.  

Having said the above, our comparative analysis together with other important recent research, poses a question as to whether $\Mcal$ should be simplified (as a result of the various levels of correlation), less subjective (weightings and transformations) and re-calibrated to better capture the elements of rural, urban inequality and ethnic minority deprivation. 

Our results also add some robust evidence to the finding of recent research on the issues surrounding the capture of deprivation in areas with large ethnic minorities \cite{Lloyd2023}.Our proposed indicator improves the identification of these areas.  

We also point out to important  feature of $\BHI$. The indicator can be easily computed and used at different levels of granularity and aggregation, be based on existing categories (MSOA, electoral wards, etc.) or any new form of clustering. Whereas this feature is also present for $\Ccal$ (albeit at output area and not as granular as postcode level) it is not the case of $\Mcal$ which is rigidly structured at LSOA level. This flexibility may of great help to research geared towards features emerging from different scaling and geographical levels. 

We also believe that the fact that the data is sourced independently from any economic and social data collection exercise to be of significant benefit for comparative analysis of performance and results, and potentially a source of quality control checks and balances. From a practical point of view, the ability to coordinate these potential checks and balances may be even easier now given that as recently as the June 2023, the sponsorship of HM Land Registry and its associated bodies moved from the Department for Business and Trade to the department of DLUHC (Department for Levelling Up, Housing and Communities). This essentially means data from the HM Land Registry and those of $\Mcal$ reside within a single overarching governance.

Lastly, we would emphasise two specific limitations to our work. Firstly, we emphasise that we made use of the word \emph{Indicator} instead of an \emph{Index} to emphasise that this is a statistically and data driven method that is subject to some level of individual mis-classifications and errors due to either local level specific dynamics or insufficient statistical data (whether $\Mcal$ should also be an indicator as opposed to an index is muted point). Indeed, the neighbourhood method described in \ref{s:neig} is precisely used to reduce data deviations. However, it is possible to construct more advance and detailed methods that may reduce uncertainty even further. Secondly, we believe that it is possible, desirable and useful to enhance the current methodology to calculate $\BHI$ by taking into account additional property information (freehold and leasehold for instance) as well as geographically specific information (location of religion sites, or council housing for instance). In this way, our research to date may be regarded as an initial point to expand the potential of embedded \emph{non monetary} information that is found within public house transactions an housing data sets as well as other public sector sources of data.

\subsection*{Ethics}

There are no specific ethical considerations to be disclosed.

\subsection*{Data Availability}

\

Data sharing not applicable to this article as no datasets were generated during the study and all data is publicly available within the terms indicated below.

\

The original data on transations, the \emph{Price Paid} data, is an open source data released under the Open Government Licence. Contains HM Land Registry data \copyright \ Crown copyright and database right 2021. This data is licensed under the Open Government Licence v3.0. \emph{Price Paid} data contains address data processed against Ordnance Survey’s AddressBase Premium product, which incorporates Royal Mail’s PAF \textregistered \ database (Address Data). Royal Mail and Ordnance Survey permit your use of Address Data in the Price Paid Data for personal and/or non-commercial use. \emph{The Office for National Statistics Postcode Directory} contains Ordinance Survey data \copyright \  Crown copyright and database right 2023.  Contains Royal Mail data \copyright \ Royal Mail copyright and Database right 2023. Contains GeoPlace data \copyright \ Local Government Information House Limited copyright and database right 2023. Source: Office for National Statistics licensed under the Open Government Licence v.3.0.


\subsection*{Author Contributions}

EV designed the overall conceptual framework, sourced the data and developed the computing coding. Both authors collaborated jointly on the development of the methodology and design, analysis and interpretation of data the writing and review of the article.

\subsection*{Competing Interests} The authors have no competing interests to declare.


\subsection*{Acknowledgments} 

The authors express their gratitude to Paul Expert from the Business School for Health at UCL on providing feedback and suggestions to the initial analysis.

\subsection*{Code Availability}

No custom code was used to develop the data processing methods described in this article. No modifications were made to the original data sourced from the relevant providers. All detailed data processing related matters were disclosed in the \emph{Methods} section and the \emph{Appendix}.



\clearpage



%
\newpage
\appendix


\renewcommand{\thesection}{\Alph{section}}
\renewcommand{\thesubsection}{\thesection.\arabic{subsection}}
\renewcommand{\thesubsubsection}{\thesubsection.\arabic{subsubsection}}

\renewcommand{\theequation}{\thesection\arabic{equation}}
\renewcommand{\thefigure}{\thesection\arabic{figure}}
\renewcommand{\thetable}{\thesection\arabic{table}}
\numberwithin{equation}{section}
\numberwithin{figure}{section}
\numberwithin{table}{section}
\setcounter{section}{0}
\renewcommand{\theHsection}{\Alph{section}}
\renewcommand{\theHequation}{\thesection\arabic{equation}}
\renewcommand{\theHfigure}{\thesection\arabic{figure}}
\renewcommand{\theHtable}{\thesection\arabic{table}}


\section{Appendix}

Additional information and background regarding the data for England and Wales.

\subsection{Buying a Property in England and Wales}

It is useful, though not essential, to understand the process of purchasing a property in England and Wales.  
This process for sales and purchases in England and Wales has some very particular and peculiar features when compared to other countries, including other parts of the United Kingdom such as Scotland. The most unusual feature is the fact that when a buyer makes an offer to purchase that is accepted by a seller, it does not constitute any legal commitment to purchase (or sell) the property. There is no early exchange of money, such as a deposit, that it is common in a number of other countries. Effectively, this means that both buyers and sellers can ``walk away'' from the transaction at any time they want, until the formal exchange of contracts is carried out. Moreover, the exchange of contracts and funds are handled by appointed solicitors holding ``client money accounts'' where banks will only provide the money from the lending on these solicitors' accounts at the completion date, which normally corresponds to the day of exchange of contract or immediately after. These mechanisms associated with the process also leads to the formation of what is commonly known as the ``sales chain'', whereby certain sellers are also buyers, and they will only complete their sale if their purchase is executed at the same time, leading to multiple sales of properties as one single group. If any of the sellers or buyers pulls out, all transactions fall through, and the sales process needs normally to start again, a situation that occurs relatively common. All the above, make the date of exchange of contracts, or the transaction date of fundamental importance, shaping the behaviour of people and professionals involved.

\subsection{Property Transaction Data}

The \href{https://landregistry-deeds.co.uk/}{HM Land Registry} is a government controlled entity that keeps all relevant property records. The registered title held there provides the evidence of the legal ownership and other matters relating to the land or property in question. Solicitors (a type of lawyer in the UK) will request information from the HM Land Registry during the sales process and once completed, all relevant sales documents are sent to be kept in custody there.

The data set from HM Land Registry on property transactions contains almost 28.5 million entries.
The data contains various errors and anomalies that leads us to exclude some of the entries. Here, we distinguish two types of exclusion. The first relates to entries where exclusion can be directly identified within the data set by specific field attributes, whereas the second type relates to exclusion that have to be identified using information beyond the orginal data set. 

With regards to the first type, we eliminated entries that (a) are simply an adjustment of the ``standard price'' paid, (b) those that relate to corrections of errors in previous entries, and (c) property types that are listed (from 2005) as not residential. The elimination of these entries reduces the data set to 27.2 million items.

For the second type of exclusion where we look outside the property transaction data we have two further refinements. Firstly, properties where the postcode within the data set could not be matched to the \href{https://geoportal.statistics.gov.uk/}{Postcode Directory} provided by the \href{https://www.ons.gov.uk/}{Office for National Statistics} were eliminated. A total of 15,827 entries (0.06\% of all entries) were removed in this way. These are most likely due to typing errors when entering the data. We were able to reduce the mismatches coming from simple variations such as the lack of spacing or double spacing. However, we did not aim to identify and correct all possible errors. Secondly, we also excluded 6,922 properties (0.02\% of all entries) where the price paid was below a threshold of \pounds 6,430. The vast majority of these exclusions on price are entries before 2003 where the categorisation of `other (non-residential) properties' was not consistently applied. For instance, these properties could be a garage for a single car on land separate from any one residential property. The threshold was set at a level where there is a clear change in the pattern of price paid amount as a function of its ordered low to high rank.

After this data cleaning, the number of property transactions used in our analysis is \emph{27,186,352} where English and Welsh transactions account for \emph{25,894,123} and \emph{1,292,229} respectively.

\subsection{Identifying Unique Properties}

In order to analyse our transaction data we need to identify any one unique property in several different transactions.  We also need to assign a precise location to this property.

In the United Kingdom every property has an address which is distinct from all other properties, allowing for deliveries to the correct property. Usually this is the number of the property along a given road, followed by the names of larger and larger geographical areas, typically suburb then city and finally county. The last element is the postcode (discussed in detail in \appref{a:postalcodes}) which is similar to the zip code in the US but which is a code shared by less than a hundred properties. The formal address is set by the Post Office, the organisation in charge of postal delivery in the United Kingdom. 

However, there is redundancy in this scheme. In principle, the house number and postal code are enough to identify a unique property but the data presents many challenges in practice.  
For a start, the data set does not contain an exclusive field for the house number, which can sometimes included within the primary line (`PAON'), the secondary line (`SAON') or within the `Street' field. 
There are many variations in the address listed in the different property transactions for the same property. 
For instance, some properties may have a name as well as a numbered location on a street or, as happens in rural areas, a house name only with no number for that road. Elements of street names may or may not be abbreviated, such as `rd.' for `road'.  One building may have several apartments which can be designated in several ways, as a separate line in the address or part of the number on the street, `Flat B, 221 Baker Street' or `221B Baker St.' and so on. Elements may be dropped e.g.\ the suburb is sometimes redundant when the name of the road is unique in that city (though this is not guaranteed to be the case for every road name). 

This lack of an identifiable unique address is reflected in the official records where the text string recording the address of a property may not always be the same, even before we take account of any errors in the recording of the address. So to identify unique properties and assign a corresponding unique ID,  $h$, the fields in the transaction data corresponding to the property address {'PAON',`SAON', `Street' and `Postcode'} must be matched as a set. 
We also note here that before applying the matching procedure we replaced any terminated postcode within the data set to the corresponding live postcodes.

The procedure results in the identification of \emph{14,838,366} unique properties, where \emph{7,416,522} properties were sold more than one time. The total transactions related to multiple sales properties is \emph{19,764,508}, or $72.7\%$ of the total transactions. From the postcode, we can assign a location to every property to an accuracy of around 100m.


\subsection{Postcodes}\label{a:postalcodes}

At the next level up, UK properties are grouped together by `\href{https://en.wikipedia.org/wiki/Postcodes_in_the_United_Kingdom}{postcodes}' which play a role similar to zip codes in the USA. The UK postcodes contain a very small number of properties, sometimes those on one side of a road in a single block but sometimes a single building with many apartments. The general format of a postcode is made up of four parts `\texttt{ab cd}' so several properties close to `221 Baker Street' share the postal code `NW1 6XE' where \texttt{a}=NW (`area code'),  \texttt{b}=1 (`district'),  \texttt{c}=6 (`sector'),  \texttt{d}=XE  (`unit'). The first part, the area code \texttt{a}, is formed by either one or two letters and corresponds to the largest geographical region the postcode. The next part, the district `\texttt{b}', is the next largest geographical region and this part always starts with a digit but can be one or two characters long. There is always a space before the second part `\texttt{cd}'. The sector `\texttt{c}' is a single numeric character and this represents a subdivision of the unit '\texttt{d}' which is two alphabetic characters. The unit can correspond to a single large building, or several buildings on part of a street or indeed on a single street. A single property is rarely defined by the postcode, and normally the house number (and perhaps the apartment number too) is necessary to identify a unique property. Currently, the are some 2.6 million postcodes within the UK for a total housing stock of around 20 million properties.

\subsection{Geographical structures}

Many counties have a strict hierarchy of geographical boundaries used for all administrative and political regions. However, the United Kingdom has many different definitions of geographical regions used for different political or administrative services. These definitions often overlap making comparison between measures defined using different definitions of areas difficult. As we noted in the main text, our method can be defined at a very fine scale so one of the main feature of the Behavioral House Indicator is that it is easily computed for any geographical structure required.

Let us consider the various definitions of geographical areas relevant to our study. At the lowest level we have the unique  postal address set by the Post Office to ensure deliveries to properties can be made reliably. At the next level up, the properties, typically no more than 100m apart, are grouped together by `\href{https://en.wikipedia.org/wiki/Postcodes_in_the_United_Kingdom}{postcodes}' as described above.

Lastly, the Office for National Statistics separates the United Kingdom regions into three distinct levels which are (from largest areas to smallest areas): the Output Area `OA', the Middle layer Super Output Area `MSOA', and the Lower layer Super Output Area `LSOA'. The boundaries for these areas are redefined after every census (carried our every ten years). Therefore, references are made to `LSOA11' , `OA21', that indicates that are the regions defined for the census 2011 and 2021, respectively.  The LSOAs tend to be the most granular level of analysis carried out by policy makers, and therefore are the basis for our research. It is possible, however, to aggregate LSOAs into distinct regions (i.e.\ local authorities, electoral wards, or parliament constituencies).


\subsection{Neighbourhood mutual information measures}\label{a:nmi}

We examined our measures for properties in specific areas. We defined a partition of the total area as set of non-overlapping smaller areas $\areaset = \{a_{1}, a_{2}, \ldots, \}$ where $a_i \cap a_j = \emptyset$ for all $i \neq j$ and $\cup_{a \in \areaset} a  = \emptyset$. 

A property can only be in one of these subdomains so we denote the set of properties in area $a \in \areaset$ as $\Hcal_{a} \subset \Hcal$ and the set of transactions of properties in area ${a}$ as $\Tcal_{a}$. 

We then define the count in area $a$ as
\begin{eqnarray}
	L_a(d_1,d_2)
	&=&
	\sum_{(h,t_i) \in \Tcal_a}
	\delta(d_1,d(t_i)) \,
	\delta(d_2,d(t_{i+1}))
	\, .
	\label{e:Ladef}
\end{eqnarray}
to match $L$ of \eqref{e:Ldef}.
This gives us the probability of a joint transaction $P_a(d_1,d_2)$ for area $a$ as 
\begin{eqnarray}
	P_a(d_1,d_2) 
	&=&
	\frac{1}{Z_a} L_a(d_1,d_2) \, , \quad d_1,d_2 \in \Wcal
	\label{e:Padoubledef}
	\\
	Z_a 
	&=& 
	\sum_{d_1 \in \Wcal} \sum_{d_2 \in \Wcal} L_a(d_1,d_2)
	\, .
\end{eqnarray}
\begin{eqnarray}
	{\Pfirst}_a(d_1) &=& \sum_{d_2\in \Wcal} \frac{L_a(d_1,d_2)}{Z_a}  = \sum_{d_2\in \Wcal} P_a(d_1,d_2)
	\, , 
	\label{e:Pafirstdef}
	\\
	{\Psecond}_a(d_2) &=& \sum_{d_1\in \Wcal} \frac{L_a(d_1,d_2)}{Z_a}  = \sum_{d_1\in \Wcal} P_a(d_1,d_2)
	\label{e:Paseconddef}
	\, .
\end{eqnarray}
Then for area $a$ we have the 
mutual information $I_a(d_1,d_2)$ and the total mutual information $W_a$ as
\begin{eqnarray}
	I_(d_1,d_2)
	&=&
	P_a(d_1,d_2) \log_{2} \left( \frac{P_a(d_1,d_2)}{{\Pfirst}_a (d_1){\Psecond}_a (d_2)}  \right)
	\, , 
	\label{e:Iadef}
	\\
	W_a 
	&=& 
	\sum_{d_1 \in \Wcal} \sum_{d_2 \in \Wcal} I_a(d_1,d_2)
	\label{e:Wadef}
\end{eqnarray}

\subsection{Local Aggregation Procedure}

\begin{figure}[htb!]
	\begin{center}
		\includegraphics[width=0.9\textwidth]{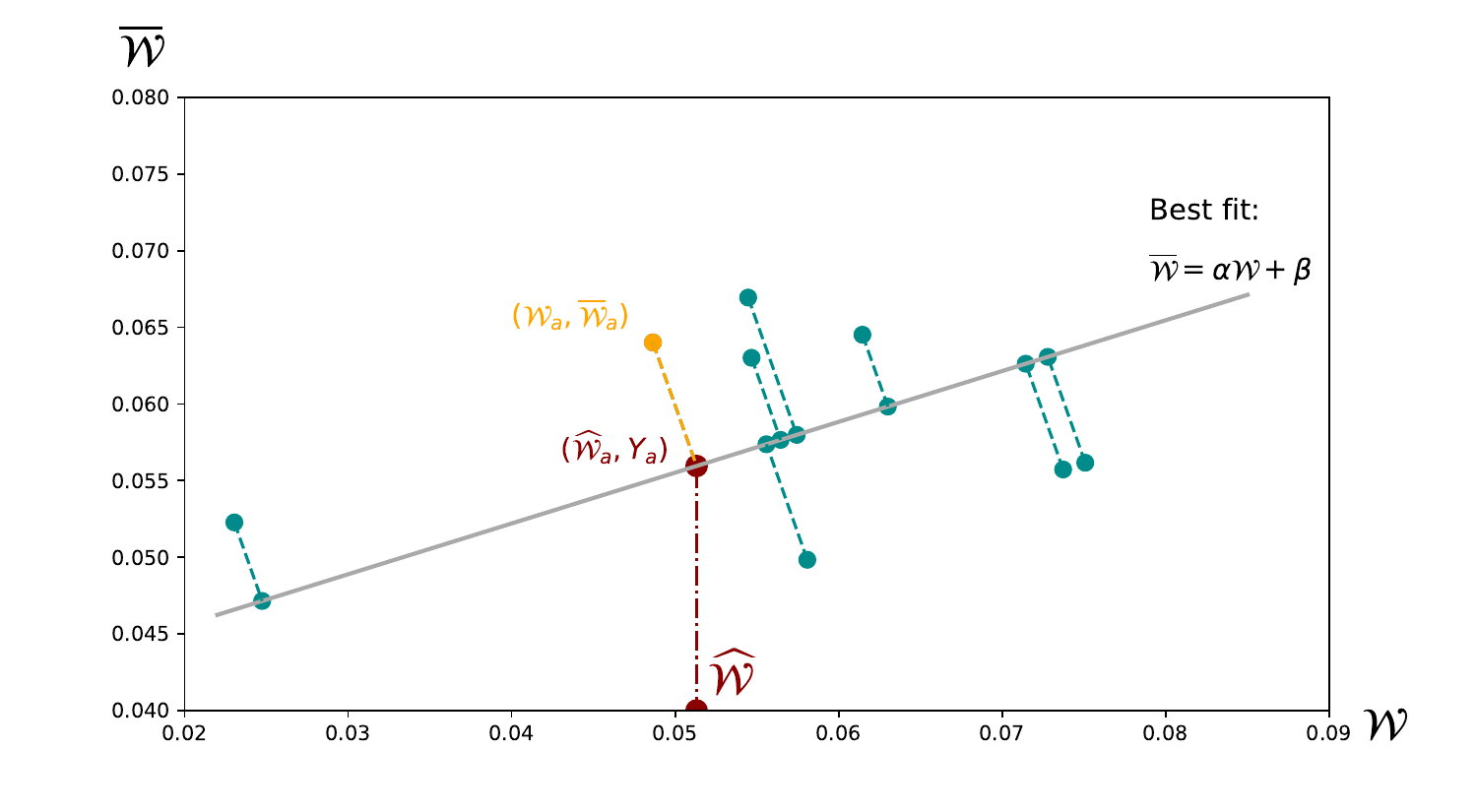}
	\end{center}
	\caption{Illustration of the method used to aggregate the mutual information $W_a$ of area ${a} \in {A}$ with the   average neighbour mutual information of its neighbours $\Wbar_a$. We find the linear fit $\Wbar = \alpha W + \beta$ to the data for all $(W_a,\Wbar_a)$ pairs. Then for each area $a$, we project onto the fitted line find the projected point $(\BHI_a ,Y_a)$. The $W$ coordinate of this projected point is then used as the aggregated mutual information value for area $a$, our Behavioural House Indicator $\BHI_a$ for area $a$. }
	\label{fig:BHIagg}
\end{figure}

In \figref{fig:BHIagg} we illustrate the method used to aggregate the mutual information $W_a$ of area ${a} \in {A}$ with the average neighbour mutual information of its neighbours $\Wbar_a$. 

\end{document}